# Detailed Thermal Characterization on a 48V Lithium-Ion Battery Pack during Charge-Discharge Cycles

Antonio Paolo Carlucci*, Hossein Darvish*, Domenico Laforgia*

Department of Engineering for Innovation, University of Salento, 73100 Lecce, Italy.

**ABSTRACT**

This study experimentally investigates the temperature distribution and behavior of a 48V Lithium-Ion (Li-ion) battery pack during two charge-discharge cycles using 25 thermocouples. Results indicate that better convective heat transfer occurs at the external surfaces of the pack, while middle cells reach maximum temperatures. Differences are also observed in the behavior of the three modules. The discharge cycle shows a temperature rise of $5.8°C$ with a pack temperature gradient increasing from $1.3°C$ to $2.7°C$. The study highlights the importance of assessing the thermal behavior of each module and the complexity of the Li-ion battery pack system. Findings on the battery cells, modules, and pack in the same study can provide valuable insights for designing efficient cooling systems for Li-ion battery packs.

**INTRODUCTION**

The Electric Vehicle (EV) is a solution to solve the problem of carbon dioxide ($CO_2$) emissions in the atmosphere. It has a lower level of pollution, and offers benefits such as lower noise, smoother operation, and regenerative braking [1]. As of mid-2022, there are about 20 million passenger EVs on the road [2]. By 2025, there will be 77 million passenger EVs on the road, representing 6% of the fleet worldwide, 13% in China and 8% in Europe [2]. Therefore, despite the short-term setbacks due to coronavirus pandemic in 2019-2022, long-term prospects for EVs remain undiminished [2].

Hybrid Electric Vehicles (HEVs) benefit from the efficiency of the EVs while providing the same range as conventional vehicles [3]. According to forecasts, HEVs are projected to capture 36% of the market by 2030 [4]. With increasing electrical system voltage levels, HEVs can be classified in micro, mild, and full HEVs [5]. Among them, 48V Mild Hybrid Electric Vehicle (MHEV) provides a promising solution for reducing both $CO_2$ emissions and cost [6,7].

Rechargeable Lithium-Ion (Li-ion) batteries are regarded as a suitable energy storage device for MHEVs. This is due to their high energy density, specific power, and light weight. In addition, they have lower self-discharge rates, higher recyclability, and longer life cycle when compared to other rechargeable batteries such as lead acid, nickel cadmium, and nickel metal hydride batteries [8-12]. Furthermore, over the last 13 years, the price of a Li-ion battery pack has dropped by almost 90% from over $1000 $kW/h$ in 2010 to $151 $kW/h$ at the end of 2022 [13,14].

Previous research on Li-ion batteries has demonstrated that high temperature and uneven temperature distribution are the main issues of the Li-ion batteries. Therefore, the temperature plays a major role in determining the life cycle and energy capacity of Li-ion batteries [15-19].

The experiments on the thermal behavior of MHEV Li-ion batteries have been conducted either on individual cells or on packs. For instance, Lin et al. investigated the thermal uniformity of two $36Ah$ pouch type Lithium Nickel Manganese Cobalt Oxide (NCM) battery cells with different sizes and under three constant charging and discharging rates [20]. They found that the temperature of the positive electrode was higher than other regions and that the current rate and direction affected the temperature gradients of the cells. Furthermore, the authors highlighted the significant influence of the battery cell size on the temperature uniformity. Recently, Kumar and Chavan examined the thermal behavior of two types of cylindrical Li-ion battery cells using numerical and experimental methods [21]. They applied three constant charging and discharging rates and various surrounding temperatures. They found out that the highest temperature rise occurred at the highest charging and discharging rates. In addition, they observed a higher temperature rise during discharge as opposed to charging at the same rate. The analysis of the open literature on this subject, however, shows that it lacks in detailed analysis of the effect of external parameters, such as the presence of other cells, on the thermal behavior of the battery cells.

---

Previous research on the thermal behavior of 48V Li-ion battery packs has been limited in scope. For instance, Lee et al. conducted an experiment on a $0.4kW$, 48V, $8Ah$ Li-ion battery pack as part of a MHEV for ground transportation [7]. They monitored the overall temperature, voltage, and current of a battery pack consisting of lithium iron phosphate cells connected in series (for a detailed explanation of how cells are connected in a battery pack, please refer to the next section, where a definition of series and parallel connections of battery cells is provided). They tested the pack during three different driving cycles and found out that the temperature of the battery pack, measured in only one single position, increased by around $6°C$ for each driving cycle. However, they did not examine the temperature distribution across different cells within the battery pack and did not provide detailed information on the battery cell dimensions and the location of the thermocouple.

In addition, Hall et al. examined the technical challenges in designing a 48V MHEV battery pack with a focus on cell selection and the thermal performance of the whole pack [22]. They used 18 lithium titanium oxide prismatic cells connected in series to make the battery pack. The cells were tested using one cycle characterized by low and high C rates at ambient temperature. They found a temperature rise of $11°C$ following a full charge or discharge event. Finally, they cycled the pack continuously charging and discharging it with a current ranging between $250A$ and $500A$, with the current switching at a frequency of 15 seconds. They found that the temperature limit of $40°C$ was reached in 8 minutes with a current of $250A$. However, they did not study the temperature distribution within the pack.

Summarizing, previous research has mainly examined Li-ion battery cells, modules, and packs with a limited number of thermocouples to monitor pack thermal behavior [6,7,20-25]. Furthermore, there is a lack of experimental studies on the detailed temperature distribution of a whole Li-ion battery pack in automotive applications.

Therefore, the present study aims to experimentally investigate the thermal behavior of a 48V Li-ion battery pack through two full charge-discharge cycles. 25 thermocouples are used at various locations within the battery pack. The high number of sensors used in this work, never adopted in open literature to the best of the authors' knowledge, is expected to provide valuable insight into the battery pack's heat generation and aid in selecting a suitable battery cooling system for the desired application in the future work.

## EXPERIMENTAL TEST

LITHIUM-ION BATTERY PARAMETERS

Basically, Li-ion battery cells consist of an anode, a cathode, an electrolyte, a separator, and current collectors. In addition, cylindrical, prismatic, and pouch shapes are employed in the automotive industry [26,27,28]. In particular, the prismatic design improves space utilization and increases flexibility [29,30].

**Table 1.** Main characteristics of battery cells and pack used in this study.

| Parameter | Value | |
|---|---|---|
| | Cell | Pack |
| Type (-) | Prismatic | 12s3p |
| Chemistry (-) | NCM | NCM |
| Nominal capacity ($Ah$) | 8.23 | 24.69 |
| Nominal voltage ($V$) | 3.7 | 44.4 |
| Maximum voltage ($V$) | 4.17 | 50 |
| Maximum current ($A$) | 233 | 700 |
| Dimensions - Thickness*Length*Width ($mm$) | 17*144*60 | 475*660*132 |

Therefore, in this study, prismatic shape battery cells are used.

In the EV battery industry, the use of NCM batteries, $LiNiMnCoO_2$, is on the rise [2,13,31]. NCM batteries allow a high C-rate in charge and discharge, a long lifespan, and good performance at low temperatures [32,33]. Therefore, in this study, NCM Li-ion battery cells are considered. The nominal capacity of the cell is rated as $8.23Ah$. The electrical and mechanical features of the cells are listed in Table 1.

This study examines the thermal performance of a 48V Li-ion battery pack shown in Figure 1a. 48V Li-ion battery pack has been chosen because it is easy to install on a vehicle, has a modular design that is safe and compact, has a low weight impact on the total vehicle weight, and has a cost-effective performance ratio in terms of fuel efficiency and torque boost.

These battery packs are made by combining one or more individual cells. The cells can be connected in parallel, series, or a combination of both to increase capacity or voltage, respectively. The battery pack being tested in this study includes 36 prismatic Li-ion cells connected in a 12s3p configuration, meaning that three cells are connected in parallel, and 12 sets of these 3-cell parallel are connected in series. The properties of the battery pack are listed in Table 1 and a detailed structure drawing of the pack, modules, and the cells is shown in Figure 1a. The cells are separated by thin thermal barriers. In the original pack, there are various wirings, a Battery Management System (BMS) circuit, electrical devices such as shunt and relay, and an aluminum chassis which have not been reproduced as they are complex both from a geometric point of view and from that of the materials used. The whole components are surrounded by a metal sheet whose dimensions are reported in Table 1.

The experimental campaign finalized to analysis of the thermal performance of the battery pack was carried out at the "Center for Studies of Vehicle Components" in Bosch plant in Modugno, Bari, Italy.

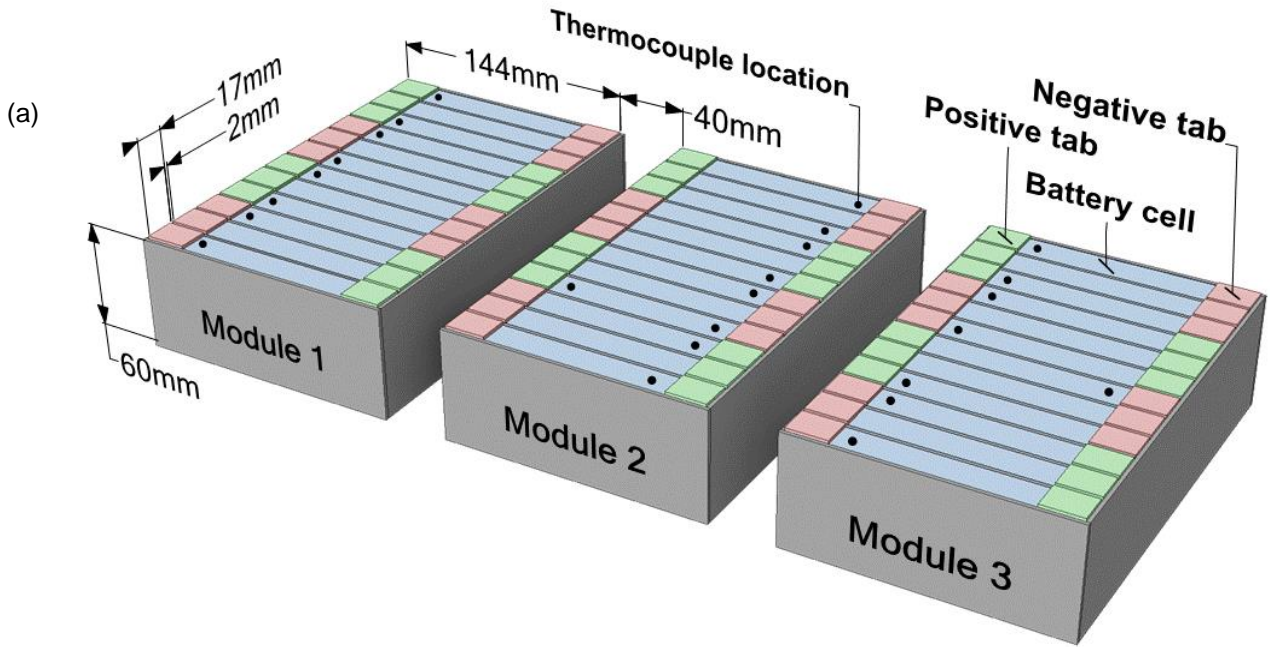

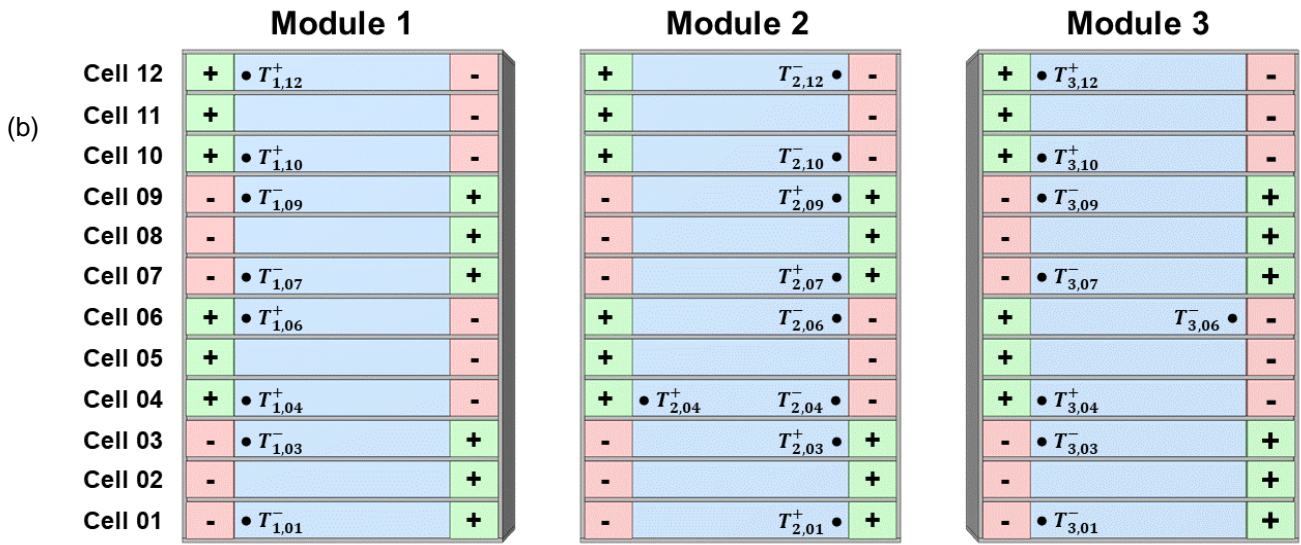

**Figure 1.** a) A trimetric view of the cells and modules to illustrate their connections, dimensions, and the positions of the thermocouples. b) A top view of the model that shows thermocouple places and names in Module1, Module2, and Module3.

EXPERIMENTAL LAYOUT

Figure 2 shows the experimental test set-up used in this study. The test bench mainly consists of four parts: a battery pack including a BMS, a high voltage - high current AVL battery emulator controlled by AVL PUMA system, K-type temperature sensors with two data acquisition modules (ES620 ETAS), and a computer unit to monitor and store the data.

As illustrated in Figure 1, 25 thermocouples are utilized to monitor the temperature of Li-ion battery cells. The various temperature measurement points are situated in three modules of the battery pack, referred to as Module1, Module2, and Module3. The naming convention for the measurement points is defined in Theoretical Background section. The highest temperature in Li-ion battery cells occurs in the vicinity of the tabs [23,24,34,35]. The tabs refer to the thin strips of metal that are attached to the electrodes within the battery cell allowing the flow of the electrons between the electrodes and the external circuit. Thus, in this study, $T^+_{2,04}$ and $T^-_{2,04}$ are placed close to the negative and positive tabs of the same cell (Figure 1b). These thermocouples help to find the variation in temperature between the positive and negative tabs of the same Li-ion battery cell during charging and discharging.

CHARGING AND DISCHARGING CYCLES

In this experiment, two full charge-discharge cycle tests have been conducted at an initial temperature and State of Charge (SoC) of $26°C$ and 47%, respectively. The maximum and minimum currents recorded during this cycle are $237A$ and $-237A$, respectively, while the SoC reaches two times the highest and lowest values of 91% and 10%. Finally, the test finishes when SoC reaches its initial value. These two cycles help to monitor the reaction of the battery pack under different SoC values in both charging and discharging loads. To avoid thermal failure of the battery pack at high temperature conditions, when the temperature of the pack reaches 40°C, the experimental test will be terminated. In this study, this limitation happens at the end of the second cycle.

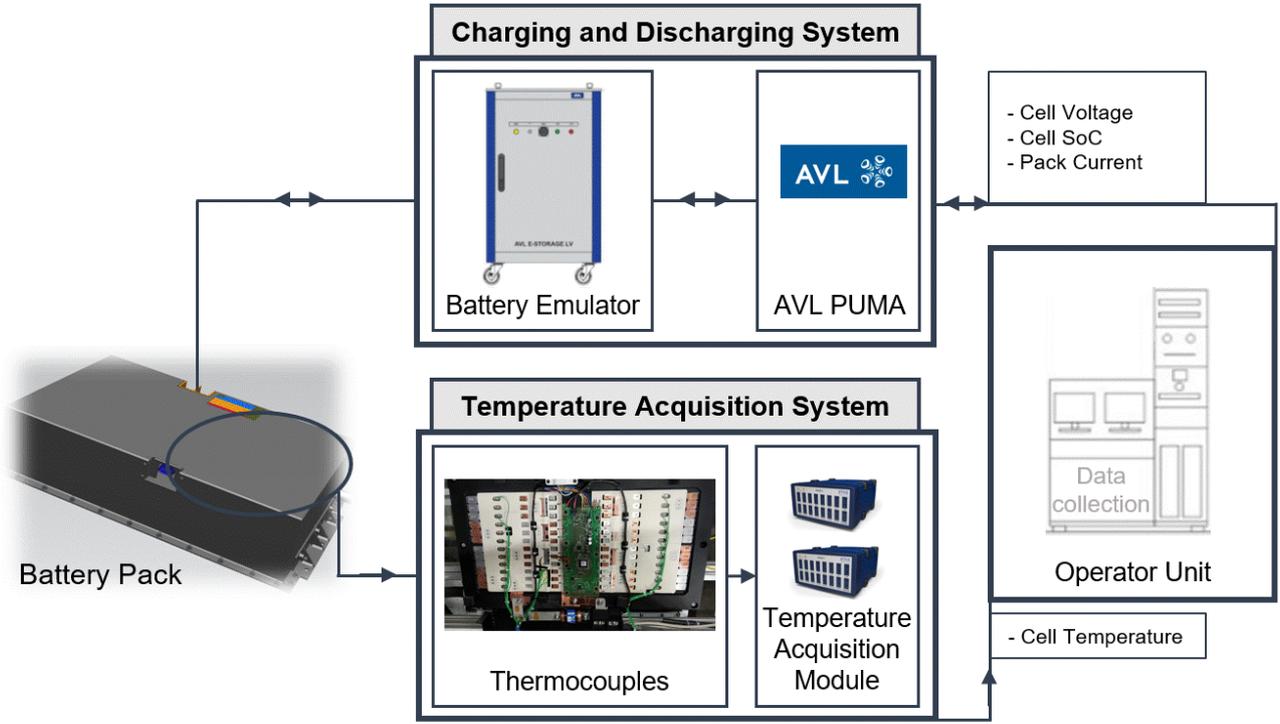

**Figure 2.** A schematic of the Li-ion battery pack thermal performance test in this study.

THEORETICAL BACKGROUND

Most of the temperature effects on batteries are related to the internal materials and chemical reactions that occur within them. Generally, heat generation within the Li-ion batteries at normal temperature ($Q_{gen}$) is associated with charge transfer and chemical reactions during charging and discharging [36,37]. The heat as in Eq. (1) is generated during either the reversible ($Q_{rev}$) or the irreversible ($Q_{irr}$) processes taking place in Li-ion batteries [36,38-40].

$$Q_{gen} = Q_{irr} + Q_{rev} \quad (1)$$

The heat generated in the reversible process, also known as entropic heat, originates from the reversible entropy change during electrochemical reactions, Eq. (2) [41]:

$$Q_{rev} = -I.T.\frac{\partial U_{OCV}}{\partial T} \quad (2)$$

where $I$ is the terminal current ($A$), whose sign is positive when discharged, $T$ is the temperature in the battery (°$C$), and $U_{OCV}$ is the open circuit voltage (V).

There are many possible irreversible processes that generate heat, including active polarization process, ohmic heating process, heating due to mixing and enthalpy change [42-44]. The irreversible heat generation can be expressed as in Eq. (3). Eq. (3) assumes that the voltage difference between the $U_{OCV}$ and terminal voltage includes a sum of all the voltage drops caused by electron and ion transport and chemical reactions. Multiplication with the current results in the irreversible heat:

$$Q_{irr} = I.(U_{OCV} - V_t) \quad (3)$$

where: $V_t$ is the terminal voltage and the open circuit voltage; $U_{OCV}$, is a function of the SoC. $Q_{irr}$ becomes zero when the current reaches zero; therefore, the heat of mixing is not considered.

According to the laws of thermodynamics, the transient behavior of the heat generated inside the battery cells produces different temperatures over time and distance. Therefore, the resulting temperature on different battery cells represents the heat generation of a complex system.

To study the thermal behavior of the Li-ion battery cell and pack, the temperature of each point is named as in Eq. (4).

$$T_{i=1,...,25}(t) = T_{m,c}^z(t) \quad (4)$$

where $m$ presents the module number and therefore is 1, 2, or 3, $c$ indicates the cell number and assumes one value in the range of 01 to 12, as each module contains 12 cells, and the superscript $z$, either - or +, represents the negative or positive tab, respectively.

The maximum and minimum temperatures ($T_{max}(t)$ and $T_{min}(t)$) are defined as the highest and lowest temperature reached in the battery module or pack at each time between the measured temperature using thermocouples. The temperature difference, $\Delta T(t)$, is calculated as the difference between $T_{max}(t)$ and $T_{min}(t)$. $T_{rise}$ is the difference between initial and final temperature in a specific period. Finally, the temperature average, $T_{avg}(t)$, is calculated as the average temperature over a specific number of temperature points.

## RESULTS

### SOC AND VOLTAGE

Figure 3 illustrates the voltage, current, and SoC of the battery pack. The current applied includes the minimum and maximum values of $237A$ and $-237A$, respectively. To have a better discussion of the results, the test time has been divided into 8 parts for the two cycles. LD, EC, LC, and ED represent Late Discharge, Early Charge, Late Charge, and Early Discharge, respectively. Numbers 1 or 2 after each of the four abbreviations respectively indicate cycle 1 or 2.

In the first section, LD1, to achieve the SoC from 47% to 10 %, the current is kept to $-237A$ for 140s. In this period, the corresponding pack and cell voltages decrease from 44.16V and 3.68V to 39.24V and 3.27V, respectively (the cell voltage is estimated as 1/12 of pack voltage). Then, in the second section, EC1, the current remains constantly equal to $237A$ until 226.5s, and SoC reaches 33% and pack voltage increases up to 44.71V.

In the third section, LC1, the current decreases to $33A$ to reach the SoC of 91%. In this period, also called the dynamic section, battery pack voltage increases up to 48.68V. At 959.5s, LC1 ends and the fourth section, ED1, starts with a constant current of $-237A$. In this section, the SoC and voltage decrease up to 47% and 42.74V, respectively. At 1120.5s, the first cycle finishes, and the second cycle starts with LD2 section.

During the second cycle, LD2, EC2, LC2, and ED2, the pack current, SoC, and voltage exhibit temporal evolution comparable to those described for LD1, EC1, LC1, and ED1. Finally, the test finishes at 2105s.

While in the literature studying Li-ion battery cells, it is common to consider only battery limitations in terms of SoC, it is worth noting that Li-ion battery pack applications in the real world often require high constant current for extended periods of time. As a result, the BMS reduces the output well below the maximum static limit to ensure safety and prevent damage. Figure 3 illustrates this trend with a constant current of 237 A, where the limit drops suddenly during the last part of each charging phase, known as LCs. The decrease in battery limits is due to battery thermal management elaborations to protect the battery.

### THERMAL PERFORMANCE

Figure 4a illustrates the temperature time histories for the 8 thermocouples placed in Module1. In the beginning, $T_{1,12}^+$ values are in the middle of the module temperature range. However, at the end of the test, $T_{1,12}^+$ decreases to the lowest. In addition, $T_{1,01}^-$ equals $T_{min}$ at the beginning of the first cycle and, generally, shows values close to the lowest recorded. The reason is their locations that are in the external surfaces of the module and have a better convective heat transfer with ambient.

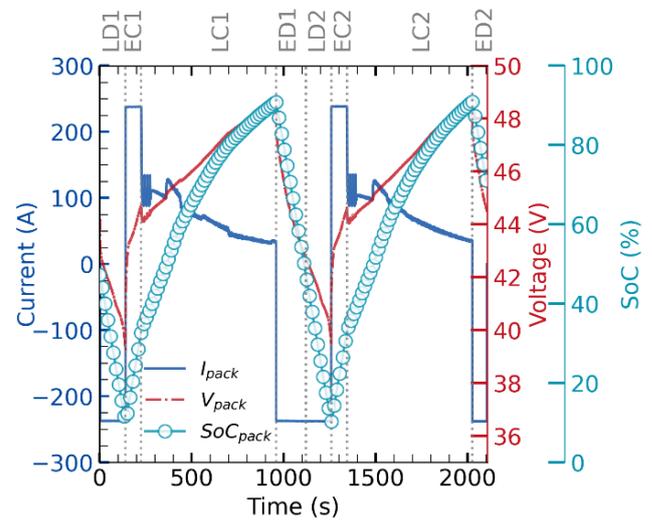

**Figure 3.** Current, SoC, and voltage of Li-ion battery pack.

Furthermore, $T_{max}$ during the test happens in the middle cells of the module, e.g., $T_{1,03}^-$, $T_{1,04}^+$, and $T_{1,06}^+$. In addition, the highest values of $T_{1,03}^-$, $T_{1,04}^+$, and $T_{1,06}^+$ for the first cycle are about 32.6°C, 32.7°C, and 32.5°C at the end of EC1 and 40.0°C, 40.2°C, and 39.9°C at the end EC2, respectively. These points exhibit a nearly constant value in the middle of LC1 and LC2 with the average temperature of 32.6°C and 36.0°C, respectively. $T_{max}$ of Module1 happens at the end of ED2 that is equal to 40.2°C in $T_{1,04}^+$.

Figure 4b shows the temperature distribution in Module2. In this module, $T_{min}$ happens in $T_{2,12}^-$. In addition, $T_{2,01}^+$ exhibits the second lowest temperature. These points are close to the boundaries of the module and therefore, there is a better cooling.

In addition, $T_{max}$ occurs in $T_{2,04}^+$. In the first cycle, it equals $31.5°C$ and $32.6°C$ at the end of EC1 and in the middle of LC1. In the second cycle, it is $39.2°C$ and $39.4°C$ at the end of EC2 and in the middle of LC2. $T_{max}$ of Module2 happens at the end of ED2 that is $39.9°C$. $T_{2,07}^+$ has almost the same temperature as $T_{2,04}^+$ in LD1, EC1, ED1, LD2, and ED2. There is a gap between these two points in LC1 and LC2 that is $0.2°C$ and $0.4°C$, respectively.

In addition, $T_{2,04}^-$ and $T_{2,04}^+$ show the temperature time histories of the thermocouples close to negative and positive tabs, respectively on the same Li-ion battery cell. $T_{2,04}^+$, the the positive tab temperature, is higher than the negative tab temperature, $T_{2,04}^-$ [23,24,34,35].

Figure 4c illustrates the temperature distributions of Module3. $T_{min}$ happens in $T_{3,01}^-$. In addition, the second lowest temperature occurs in $T_{3,12}^+$. This is also because of a better conductive heat transfer with the boundary at these points. Furthermore, $T_{max}$ for Module3 happens in three internal cells, $T_{3,04}^+$, $T_{3,06}^-$, and $T_{3,07}^-$. They are equal to 31.6°C, 32.0°C, 31.5°C, at the end of EC1, 39.1°C, 39.5°C, 39.1°C, at the end of EC2, 32.5°C, 32.5°C, 32.5°C, in the middle of LC1, and 38.9°C, 38.8°C, 39.1°C, in the middle of LC2, respectively. Between the mentioned three high temperature cells, $T_{3,06}^-$ in LD1, EC1, LD2, and EC2, is higher than $T^+_{3,04}$ and $T^-_{3,07}$ in addition to the beginning of LC1 and LC2. The maximum difference happens at the end of EC1 and EC2. On the other hand, in the remaining time in LC1 and LC2, the lowest current time history is lower than the other two cell temperatures. Therefore, in high current, there is a difference between $T^-_{3,06}$, the warmest cell in Module3, and the second high temperature cell. On the other side, the temperature of the two other points is higher than the first one in low current charging.

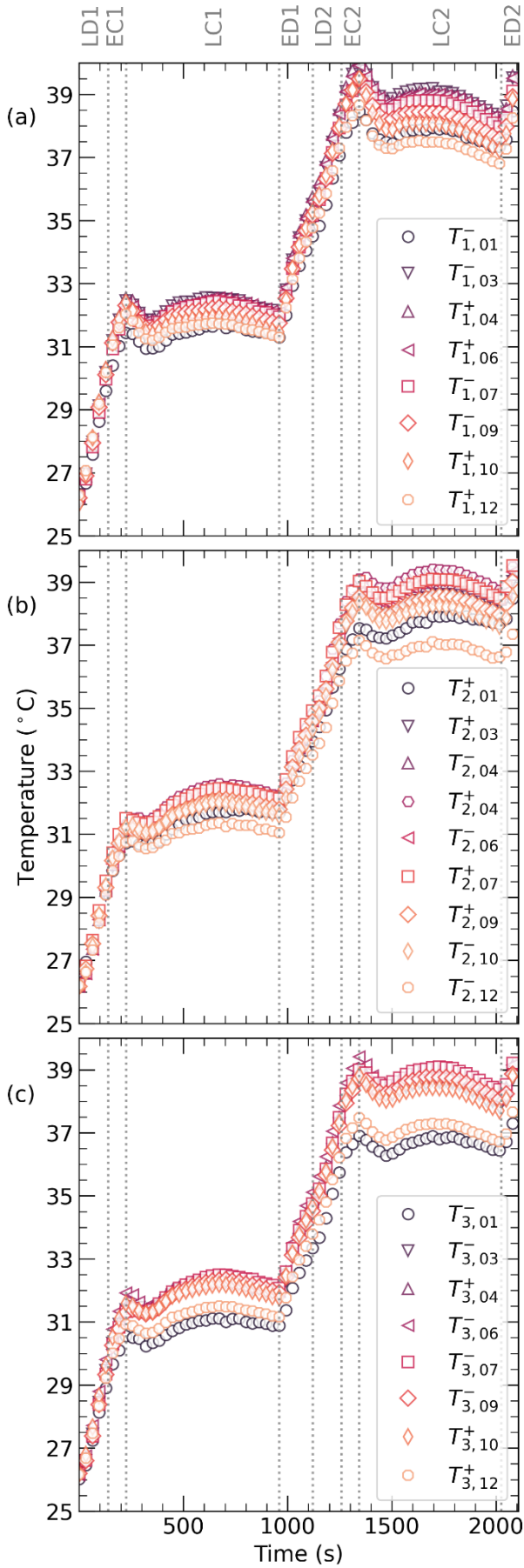

**Figure 4.** Temperature time histories of the thermocouples placed in a) Module1, b) Module2, and c) Module3.

Figure 5 shows the time histories related to: $T_{avg}$ of the whole battery pack and for each module; $T_{max}$ and $T_{min}$; $\Delta T$ of the pack. $T_{avg}$ of the pack at LD1, EC1, LC1, ED1, LD2, EC2, LC2, and ED2 are 29.9°C, 31.5°C, 31.8°C, 34.7°C, 37.6°C, 38.8°C, 37.8°C, and 39.3°C, respectively. Therefore, the minimum and maximum $T_{avg}$ occur in LD1 and ED2, respectively.

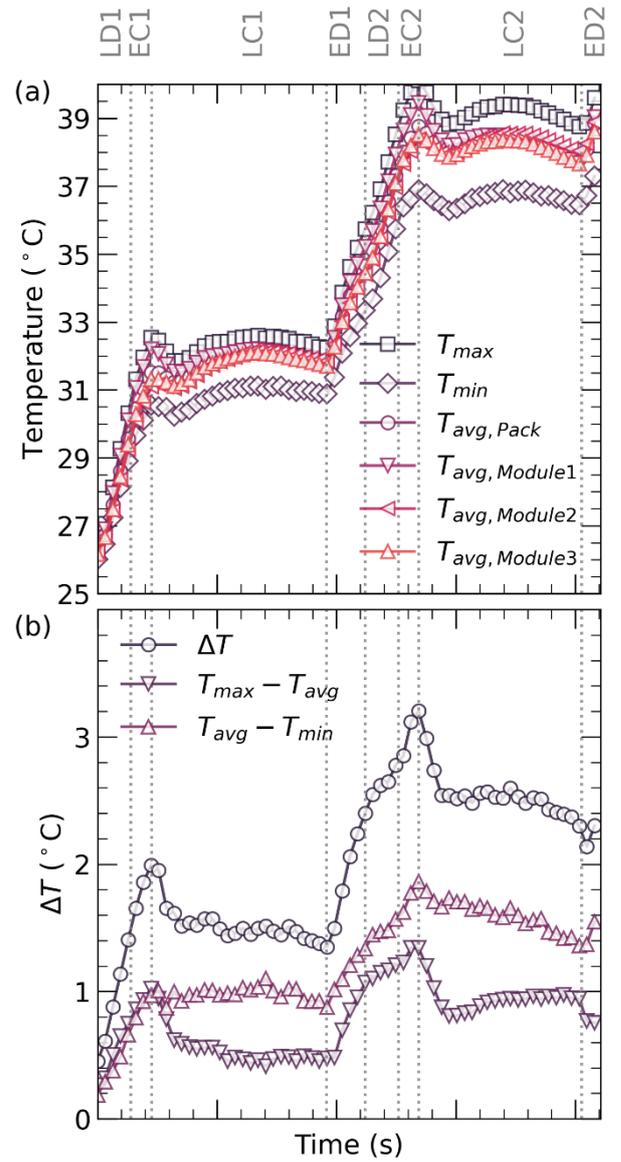

**Figure 5.** a) $T_{avg}$ of the pack and modules, as well as $T_{max}$, $T_{min}$ of the pack; b) pack $\Delta T$ behavior and its components.

$T_{avg}$ of Module1 behaves in the same way of $T_{avg}$ for the whole battery pack at the end of LC1 and LC2. In the remaining time periods, it is always higher than $T_{avg}$ of the pack. For instance, $T_{avg}$ of Module1 at the end of LD1, EC1, ED1, LD2, EC2, and ED2, is 0.6°C, 0.7°C, 0.5°C, 0.6°C, 0.7°C, and 0.3°C higher than $T_{avg}$ of the pack. Therefore, Module1 is more sensitive to the current changes and there is a difference between its $T_{avg}$ and the two other modules. $T_{avg}$ of Module2 and Module3 behave almost the same in respect to the pattern of the rising temperature and its value. It shows the importance of tracking all the module's temperature.

$T_{rise}$ of the pack is 1.6°$C$ and 1.2°$C$ in EC1 and EC2, respectively. As shown in Figure 5, the profiles of $T_{avg}$ in LC1 and LC2 are almost the same. First, they decrease and then with a low gradient they increase and finally they decrease again till starting the next discharge procedure. Therefore, $T_{avg}$ of the pack in the second half of the charging procedures, in both cycles, remains quasi steady. On the other hand, $T_{rise}$ in the full discharge cycle (the combination of ED1 and LD2) is 5.8°$C$ in about 300 seconds.

The maximum $\Delta T$'s shown in Figure 5 happen at the end of EC1 and EC2, equal to 2.0°$C$ and 3.2°$C$, respectively. In addition, the minimum $\Delta T$ occurs at the end of the first and second full charge cycles, the end of LC1 and LC2. They are equal to 1.3°$C$ and 2.2°$C$. In the first and second full charge cycles, $\Delta T$ increases in the ECs. It is because of a high current of 237$A$. Then, it decreases and in the second half of the LCs, $\Delta T$ is almost constant. In addition, in the full discharge cycle (the combination of ED1 and LD2) $\Delta T$ increases constantly from 1.3°$C$ to 3.2°$C$.

Finally, $\Delta T$ can be divided into two parts as illustrated in Figure 5. The difference between $T_{max}$ and $T_{avg}$, and the difference between $T_{avg}$ and $T_{min}$. The difference between $T_{avg}$ and $T_{min}$ changes linearly but when the current changes significantly, the slope of the line changes. The difference between $T_{max}$ and $T_{avg}$ is much sensitive to the current and its behavior is not linear. In this case, in addition to the abrupt changes over the significant current change, over a constant and dynamic current, a dynamic behavior of the temperature difference is seen.

## DISCUSSION

### CELL STUDY

On the same Li-ion battery cell, the temperature of the positive tab is higher than the temperature of the negative tab, as shown by $T_{2,04}^-$ and $T_{2,04}^+$ showing the temperature of the points close to the negative and positive tabs, respectively. The maximum difference is about 0.6°$C$ and this phenomenon was seen in the literature as well [23,24,34,35].

In addition, $T_{min}$ at the end of two cycles happens in $T_{1,12}^+$ and $T_{1,01}^-$, $T_{2,12}^-$ and $T_{2,01}^+$, and $T_{3,01}^-$ and $T_{3,12}^+$ in Module1, Module2, and Module3, respectively. Therefore, $T_{min}$ happens on the external cells of the battery pack. This is due to a better convective heat transfer and therefore, a better cooling in respect to ambient temperature at the module boundary surfaces. On the other hand, $T_{max}$ of each module happens in middle cells of the module. However, this is not symmetric and shows the dynamic behavior of each cell through the time and then the temperature ununiformly in a Li-ion battery pack. This phenomenon illustrates the complexity of such a dynamic system and highlights the importance of the evaluation of all the modules' temperature behavior in a battery pack.

### MODULE STUDY

$T_{avg}$ of Module1 is higher than pack $T_{avg}$ in LDs, ECs, EDs, and the first half of LCs. It shows that Module1 is more sensitive to high current than other modules and produces more heat. In addition, its temperature increases faster than the whole pack and as a result, it also exchanges heat better than others. It shows the complexity of the battery pack system, and that the thermal behavior of each module should be studied and examined separately.

### PACK STUDY

In the Li-ion battery cell, module, and pack, the temperature always rises in EDs, LDs, and ECs. Therefore, $T_{max}$ over the time happens at the end of EC1 and EC2 in addition to the middle of the LC1 and LC2. In other words, the temperature rises when there is high current as illustrated in Eq. (2). This is because of a higher number of Li-ions that should move through the separator and produce more heat. Therefore, at the beginning of LCs, a temperature drop happens and following that, a quasi-steady state behavior of the temperature is observed.

In the full discharge cycle, between the two charging cycles, the temperature rises monotonically; the overall increase, $T_{rise}$, is 5.8°$C$ from an initial value of 31.8°$C$. In addition, the same rising behavior happens for the $\Delta T$ from 1.3°$C$ to 2.7°$C$. This is because of the high current at this period based on Eq. (2), and the sensibility of Module1 to the current. Furthermore, $T_{rise}$ pattern in the two charging cycles, show similar trends. In the beginning, it rises, and then decreases, and finally the quasi-steady temperature remains. Therefore, when a higher current is applied on the pack, a higher $T_{rise}$ and a higher $\Delta T$ of the pack are reached.

As illustrated in Figure 5, the $\Delta T$ can be divided into two parts: the gap between $T_{max}$ and $T_{avg}$ and the gap between $T_{avg}$ and $T_{min}$. $T_{max}$ is much sensitive to the temperature gradient and changes through the time, while $T_{min}$ is less sensitive to the current changes. Therefore, the most important part of the temperature changes of the battery pack is because of $T_{max}$ behavior. In other words, the difference between $T_{avg}$ and $T_{min}$ changes linearly but when there is a large difference in the current, the slope changes. The difference between $T_{max}$ and $T_{avg}$ is much sensitive to the current and over the time there are higher slopes. Therefore, in the second case, in addition to the big changes over a large current difference, over a constant and dynamic current, there is not a linear behavior of $\Delta T$.

Future work should focus on developing efficient cooling systems and exploring different thermal management strategies to enhance the performance and safety of Li-ion battery packs based on the results obtained in this study.

## Conclusion

In conclusion, this study investigated the thermal behavior of a 48V Li-ion battery pack during dynamic current profiles. This study is important because understanding the thermal behavior of Li-ion battery packs is crucial for their safe and reliable operation, especially in applications where high power and energy densities are required.

The results obtained from the experimental tests showed that the temperature behavior of the battery pack is complex and nonlinear, and it varies across different cells, modules, and the entire pack. The temperature of the external cells of the battery pack is lower than the internal cells due to better convective heat transfer. Moreover, the positive tab temperature is higher than the negative tab temperature on a single cell. One module is more sensitive to high current; therefore, it leads to a faster temperature rise and heat generation. This specific result could only be reached by studying all the modules in the same Li-ion battery pack.

The temperature behavior of the battery pack is primarily influenced by $T_{max}$, which is much more sensitive to the current compared to $T_{avg}$ and $T_{min}$. The $T_{rise}$ is mainly due to high current and the heat generated by moving Li-ions through the separator. The difference between $T_{min}$ and $T_{avg}$ changes linearly over a constant current and the difference between $T_{max}$ and $T_{avg}$ changes nonlinearly with the current, especially over a large current difference.

Overall, the findings suggest the importance of examining and evaluating the thermal behavior of each cell, module, and the entire pack separately to understand the complexity and nonlinear behavior of Li-ion battery packs during dynamic current profiles. The results of this study could contribute to the future work to develop more efficient and reliable battery thermal management systems for Li-ion battery packs in automotive applications. In particular, the difference between $T_{avg}$ of modules, dependency of $T_{avg}$ to $T_{max}$, and the impact of the cell location on its temperature changes have to be considered in determining battery thermal management strategies and parameters.

## Credit Authorship Contribution Statement

**Antonio Paolo Carlucci:** Conceptualization, Validation, Resources, Writing – review & editing, Supervision, Project administration. **Hossein Darvish:** Conceptualization, Methodology, Software, Validation, Formal analysis, Investigation, Resources, Data curation, Writing – original draft, Writing – review & editing, Visualization, Project administration. **Domenico Laforgia:** Validation, Resources, Supervision, Project administration, Funding acquisition.

## CONTACT

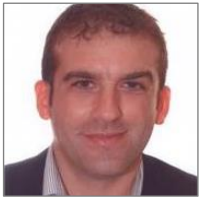

**Antonio Paolo Carlucci** is an associate professor in the Department of Engineering for Innovation at the University of Salento, Lecce, Italy. He is currently the director of "Engines, Combustion and Spray" lab. where research activities on combustion using liquid and gaseous alternative and renewable fuels in dual-fuel compression ignition and spark ignition engines are carried out. He has been involved in national and international projects on renewable fuels exploitation, energy recovery and energy production integrated systems. He coauthored numerous journal and conference papers and four book chapters.

✉ paolo.carlucci@unisalento.it

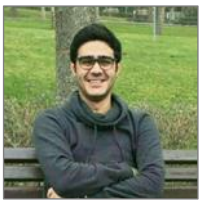

**Hossein Darvish** is a Ph.D. candidate in the Department of Engineering for Innovation at the University of Salento, Lecce, Italy. His research focuses on Thermal-Fluid Transport projects with applications to Lithium-Ion (Li-ion) Battery Thermal Management Systems (BTMS) and Thermal Energy Storage, as well as Cardiovascular Biomechanics. Currently, he is completing his Ph.D. thesis at the "Center for Studies of Vehicle Components" in Bosch plant in Modugno, Bari, Italy, where he is involved in analyzing 48V Li-ion battery packs and developing innovative climate chamber test benches to optimize BTMS.

✉ hossein.darvish@unisalento.it

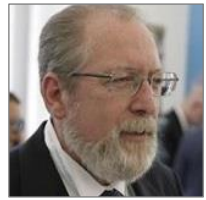

**Domenico Laforgia** is an Emeritus Professor in Energy Systems and Fluid Machinery and Scientific Coordinator of CREA research group at the University of Salento, Lecce, Italy. He coordinated as Scientific Responsible of many national and international projects on energy efficiency, renewable fuels exploitation, energy recovery and energy production integrated systems. He has published over 350 international scientific articles and 17 books as well as seven patents in the field of energy efficiency.

✉ domenico.laforgia@unisalento.it

## ACRONYMS

**BMS:** Battery Management System

**EC:** Early Charge

**ED:** Early Discharge

**EV**: Electric Vehicle

**HEV**: Hybrid Electric Vehicle

**LC:** Late Charge

**LD:** Late Discharge

**MHEV**: Mild Hybrid Electric Vehicle

**NCM**: Lithium Nickel Manganese Cobalt Oxide

**Li-ion**: Lithium-Ion

**SoC**: State of Charge

**Abstract (200-300):**

This study experimentally investigates the thermal behavior of a 48V Lithium-Ion (Li-ion) battery pack during two full charge-discharge cycles. The battery pack consists of three identical modules, each containing 12 prismatic nickel manganese cobalt oxide battery cells. Using 25 thermocouples, the temperature distribution of the battery pack is measured and analyzed to determine temperature changes in the cells, modules, and pack. Results show that the minimum temperature occurs at the external surfaces of the battery pack due to better convective heat transfer, while the maximum temperature occurs in the middle cells of each module. One module is found to be more sensitive to high current than the other modules, producing more heat and releasing it faster. Moreover, the positive tab temperature is higher than the negative tab temperature on a single cell. The pack temperature rise during the discharge cycle with a constant current of -237A is found to be 5.8°C from 31.8°C that is mainly due to high current and the heat generated by moving Li-ions through the separator, while the same rising behavior is observed for the pack temperature gradient from 1.3°C to 2.7°C. Finally, the difference between minimum and average pack temperatures changes linearly over a constant current and the difference between maximum and average temperatures changes nonlinearly, especially over a large current difference. This study highlights the importance of evaluating the thermal behavior of each module separately and the complexity of the Li-ion battery pack system. The findings on the battery cells, modules, and pack in the same study can provide valuable insights for designing efficient cooling systems for Li-ion battery packs.